\documentclass[conference]{IEEEtran}
\usepackage{latexsym}
\usepackage{float}
\usepackage{cite}
\usepackage{amsmath,amssymb,amsfonts}
\usepackage{algorithmic}
\usepackage{graphicx}
\usepackage{textcomp}
\usepackage{xcolor}
\usepackage{amsfonts}
\usepackage{amsbsy}
\usepackage{amssymb}
\usepackage{times}
\usepackage{graphicx}
\usepackage[strings]{underscore}
\usepackage{enumerate}
\usepackage[dvips]{pstcol}
\usepackage{epstopdf}
\usepackage{cite}
\usepackage{amssymb}
\usepackage{amsfonts}
\usepackage{graphicx}
\usepackage{epsfig}
\usepackage{psfrag}
\usepackage{xcolor}
\usepackage{amsfonts, bm}
\usepackage{epstopdf}
\usepackage{cite}
\usepackage{color}
\usepackage{xcolor}
\usepackage{subfig}
\usepackage{verbatim}
\usepackage{multirow}
\usepackage{array}
\usepackage{booktabs}
\usepackage{amsthm}
\usepackage{makecell}
\usepackage{subfig}

\IEEEoverridecommandlockouts

\begin{document}
	\vspace{-3.3em}
	\IEEEpeerreviewmaketitle
	\title{Deep Refinement-Based Joint Source Channel Coding over Time-Varying Channels \\
	}
	
	\author{\IEEEauthorblockN{ Junyu Pan, Hanlei Li, Guangyi Zhang,  Yunlong Cai, and Guanding Yu   }
		\IEEEauthorblockA{ College of Information Science and Electronic Engineering, Zhejiang University, Hangzhou, China }
		E-mail: \{junyupan, hanleili, zhangguangyi, ylcai, yuguanding\}@zju.edu.cn} 
	
	\maketitle
	\begin{abstract}
		In recent developments, deep learning (DL)-based joint source-channel coding (JSCC) for wireless image transmission has made significant strides in performance enhancement.
		Nonetheless, the majority of existing DL-based JSCC methods are tailored for scenarios featuring stable channel conditions, notably a fixed signal-to-noise ratio (SNR).
		This specialization poses a limitation, as their performance tends to wane in practical scenarios marked by highly dynamic channels, given that a fixed SNR inadequately represents the dynamic nature of such channels.
		In response to this challenge, we introduce a novel solution, namely deep refinement-based JSCC (DRJSCC). This innovative method is designed to seamlessly adapt to channels exhibiting temporal variations.
		By leveraging instantaneous channel state information (CSI), we dynamically optimize the encoding strategy through re-encoding the channel symbols. This dynamic adjustment ensures that the encoding strategy consistently aligns with the varying channel conditions during the transmission process.
		Specifically, our approach begins with the division of encoded symbols into multiple blocks, which are transmitted progressively to the receiver. 
		In the event of changing channel conditions, we propose a mechanism to re-encode the remaining blocks, allowing them to adapt to the current channel conditions.  
		Experimental results show that the DRJSCC scheme achieves comparable performance to the other mainstream DL-based JSCC models in stable channel conditions,
		and also exhibits great robustness against time-varying channels.
		
	\end{abstract}

	\section{Introduction}
	In line with Shannon's separation theorem \cite{Shannon}, the prevailing approach in modern communication systems begins with source coding, such as JPEG2000 \cite{JPEG}, to compress the source data, followed by the application of a source-independent channel code, like Turbo code \cite{Turbo}, for error protection.
	However, viewed through the lens of information theory, this separate approach is constrained in achieving optimal performance within the practical finite blocklength regime. 
	Conversely, the joint source-channel coding (JSCC) scheme, which outperforms the separate approach in finite blocklength scenarios \cite{JSCC_better}, faces limited adoption due to the intricate design requirements of effective JSCC schemes.
	
	Recently, the evolution of deep learning (DL) has spurred researchers to leverage deep neural networks (DNNs) in the parameterization of JSCC schemes, resulting in the development of effective systems with manageable complexity.
	This DL-based JSCC approach has demonstrated superiority over traditional separate schemes \cite{compare} and has found applications in the transmission of multimodal data encompassing text, speech, image, and video \cite{DeepSC,speech,video,DeepJSCC,Deep_F,UDeepSC,ADJSCC,Deep_V}.
	For instance, the authors of \cite{DeepSC} introduced the DeepSC framework, based on the Transformer, for text transmission.
	DL-based JSCC schemes have also been proposed to directly map signals to channel symbols in scenarios involving speech \cite{speech} and video \cite{video} transmission. 
	In the realm of wireless image transmission, the authors of \cite{DeepJSCC} presented a DL-based JSCC scheme, known as deep JSCC, which outperformed the separate scheme, particularly in conditions marked by low signal-to-noise ratio (SNR).
	In reference to \cite{DeepJSCC}, the approach presented in \cite{Deep_F} harnessed channel output feedback, resulting in a remarkable enhancement in end-to-end reconstruction quality for transmissions of fixed length. 
	Despite the notable achievements of these strategies, their primary focus lies in optimizing models within a fixed SNR regime. 
	This approach can result in performance degradation due to the inherent mismatch between the training phase and the deployment phase. To tackle this issue, several methods have been proposed \cite{ADJSCC,Deep_V}.
	The attention DL-based JSCC (ADJSCC) model introduced in \cite{ADJSCC} can be trained across a range of SNRs by incorporating attention feature (AF) modules and taking SNR as an input. 
	This approach significantly improves performance and demonstrates increased resilience to SNR mismatch compared to existing models.
	Additionally, in \cite{Deep_V}, authors proposed a predictive and adaptive deep coding framework designed to achieve flexible code rate optimization while meeting specific transmission quality requirements. 
	This model performs on par with ADJSCC in terms of achieving similar performance outcomes.
	
	The schemes mentioned above primarily cater to a straightforward channel model, such as additive white Gaussian noise (AWGN). They assume a constant channel state throughout the transmission of a single image. In this context, the coding strategy for each image is tailored for a fixed SNR, which gives rise to two primary issues:
	\begin{itemize}
		
		\item In real-world scenarios, both the length of image codes and channel coherence time are subject to uncertainty. Channel conditions may exhibit variations within the transmission of a single image.

		\item Performance deteriorates rapidly when there is a disparity between the actual SNR during deployment and the SNR considered during the design stage.
		
	\end{itemize}
	As a result, a fixed SNR falls short in encapsulating the dynamic nature of channel conditions, thus imposing constraints on the real-world deployment of DL-based JSCC solutions.

	\begin{figure*}[t]
		\centering 
		\includegraphics[width=0.75\linewidth]{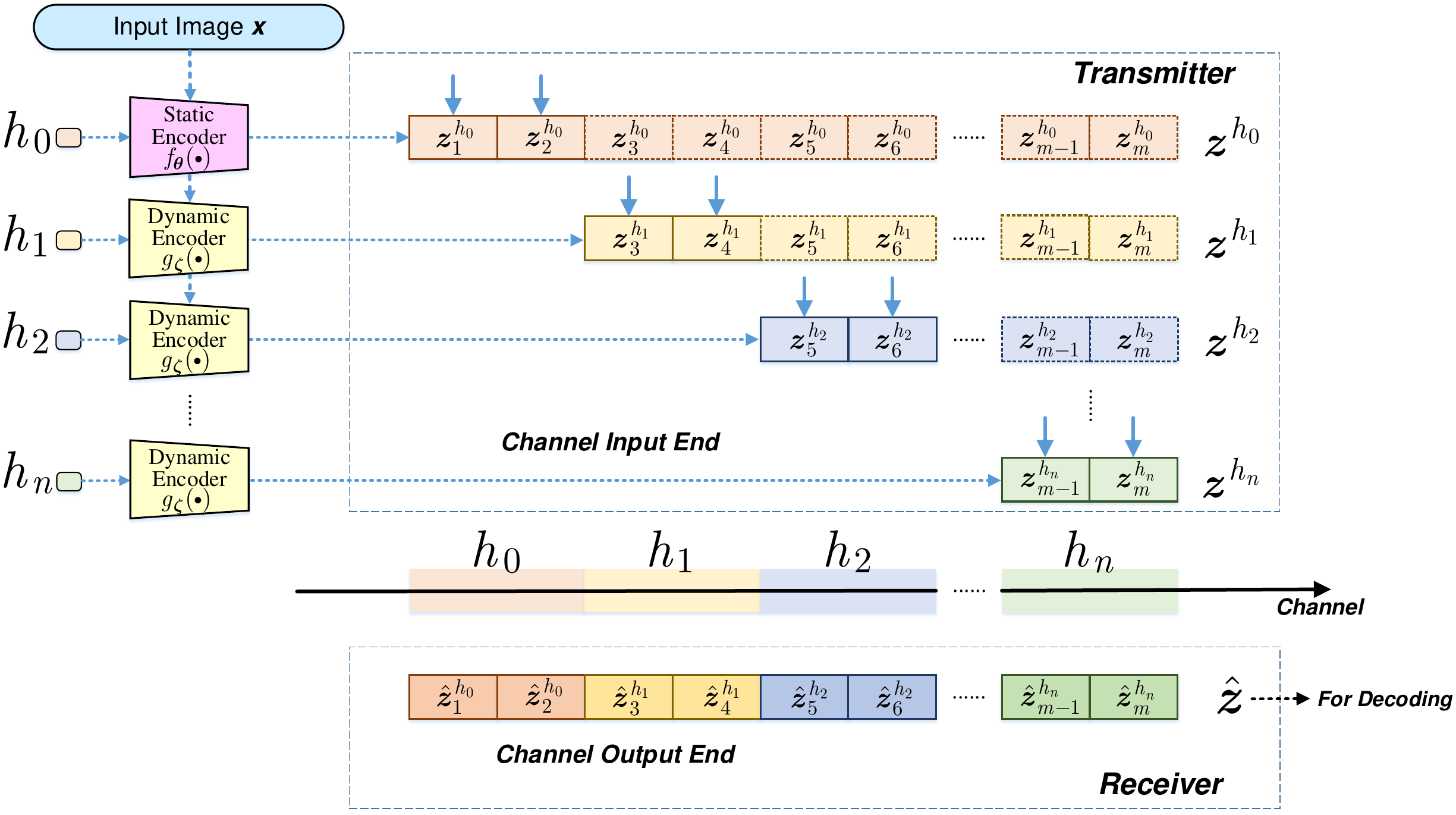} 
		\captionsetup{font={footnotesize  }}
		\captionsetup{justification=raggedright,singlelinecheck=false}
		\caption{The process of encoding and transmission.}
		\label{fig:transmission} 
	\end{figure*}
	To  tackle  this issue, we propose a novel DL-based JSCC method known as deep refinement-based JSCC (DRJSCC).
	DRJSCC possesses the ability to dynamically refine the encoding strategy by re-encoding channel symbols, utilizing real-time feedback on channel state information (CSI). 
	Specifically, DRJSCC partitions encoded symbols into multiple blocks, transmitting them progressively to the receiver. 
	When channel conditions undergo changes, DRJSCC promptly re-encodes the remaining blocks to align with the current channel state.
	This dynamic approach endows DRJSCC with notable advantages, including superior robustness compared to alternative methods and increased adaptability to address the fluctuations associated with time-varying channel conditions. These variations may encompass different image code lengths and changes in channel coherence time.
	Simulation results demonstrate that our proposed DRJSCC achieves performance metrics nearly on par with existing DL-based JSCC models in stable channel conditions,
	but with the capacity to adeptly navigate the challenges posed by dynamic and time-varying channels.

	\section{Implementation of Refinement Mechanism} 
	We consider an end-to-end image transmission system. 
	In this setup, the encoder at the transmitter transforms the source value, denoted as $\bm{x}$, into channel input symbols, referred to as $\bm{z}$. These input symbols are subsequently conveyed through a channel characterized by noise, ultimately yielding channel output symbols, represented as $\hat{\bm{z}}$.
	The decoder at the receiver utilizes $\hat{\bm{z}}$ to obtain the reconstructed image $\hat{\bm{x}}$. Our emphasis is on the Rayleigh fading channel. The channel
	transmission process is mathematically represented as follows:
	\begin{equation}
		\hat{\bm{z}}=h\bm{z}+\bm{n},
	\end{equation}
	where $h \in \mathbb{C}$ denotes the channel gain coefficient, and 
	$\bm{n} \sim \mathcal{CN}(0,\sigma^2\bm{I})$ denotes AWGN.
	
	It is important to note that, in contrast to prior studies that assume a single image experiences an identical fading event, our DRJSCC framework addresses situations where images may encounter several distinct fading events. To this end, our proposed DRJSCC consists of the static encoder/decoder and the dynamic encoder/decoder. 
	At the outset, the channel gain coefficient is denoted as $h_0$.  
	We operate under the assumption that this channel gain coefficient can be estimated by the receiver and subsequently fed back to the transmitter.
	An input image of size $C$ (channel) $\times$ $H$ (height) $\times$ $W$ (width) is represented by a vector $\bm{x} \in \mathbb{R}^{l}$, where $l=C\times H \times W$.
	The static encoder performs encoding on $\bm{x}$ and $h_0$ to a set of complex-valued channel symbols $\bm{z}^{h_0}$ and divides $\bm{z}^{h_0}$ into $m$ blocks, which can be expressed as:
	\begin{equation}
		\bm{z}^{h_0} =\{{\bm{z}^{h_0}_i}\}^m_{i=1}= f_{\bm{\theta}}(\bm{x},h_0) \in \mathbb{C}^k,
	\end{equation}
	where $k$ denotes the size of channel input symbols, $\bm{z}^{h_0}_i \in \mathbb{C}^{k/m}$ denotes the $i$-th block of $\bm{z}^{h_0}$,
	$f$ and $\bm{\theta}$ denote the static encoder and its parameter set, respectively.
	Additionally, $\bm{z}^{h_0}$ is subject to a power
	constraint $P$, i.e., $\frac{1}{k}  \mathbb{E} ||\bm{z}^{h_0}||^2 \leq P$.
	
	\begin{figure*}[h]
		\centering 
		\includegraphics[width=0.73\linewidth]{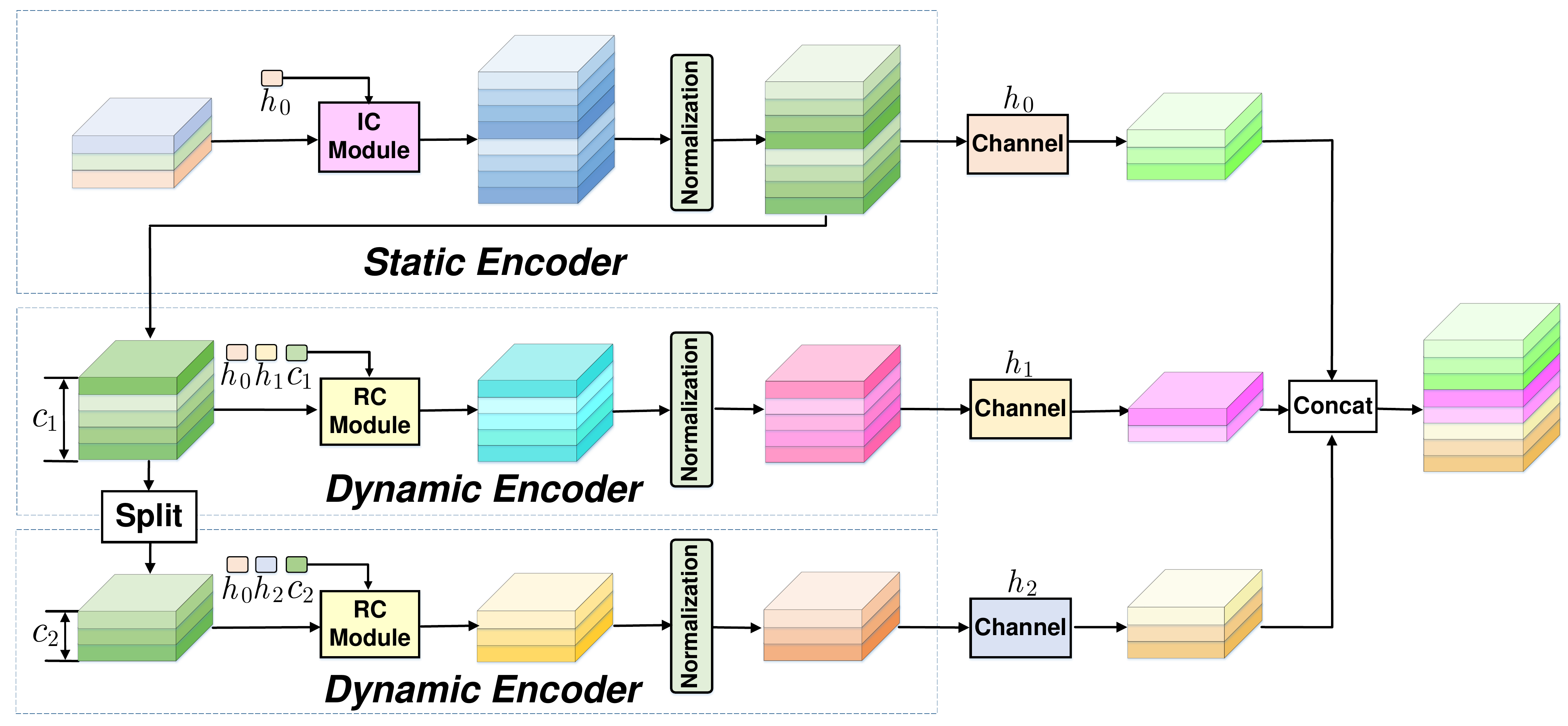} 
		\captionsetup{font={footnotesize  }}
		\captionsetup{justification=raggedright,singlelinecheck=false}
		\caption{The encoders of the proposed DRJSCC.}
		\label{fig:system} 
	\end{figure*}
	When the channel gain coefficient remains constant, the blocks are transmitted in sequence and stored at the channel output end. However, when the channel gain coefficient undergoes changes, the dynamic encoder re-encodes the remaining blocks.
	We consider a simplified scenario as shown in Fig. \ref{fig:transmission} as an example for further illustration.
	
	\begin{enumerate}[(i)] 
		\item After $[\bm{z}^{h_0}_1,\bm{z}^{h_0}_2]$ are  transmitted under $h_0$, The channel gain coefficient changes to $h_1$. 
		The dynamic encoder refines the encoding strategy with $h_1$ and re-encodes $(\{{\bm{z}^{h_0}_i}\}^m_{i=3})$ to $(\{{\bm{z}^{h_1}_i}\}^m_{i=3})$ for the next transmission.
		
		\item After $[\bm{z}^{h_1}_3,\bm{z}^{h_1}_4]$ are  transmitted under $h_1$, The channel gain coefficient changes to $h_2$. 
		The dynamic encoder refines the encoding strategy with $h_2$ and re-encodes $(\{{\bm{z}^{h_0}_i}\}^m_{i=5})$ to $(\{{\bm{z}^{h_2}_i}\}^m_{i=5})$ for the next transmission.
		
		\item This process repeats until all blocks have been progressively transmitted to the receiver in multiple steps.
		The blocks are concatenated to form the final channel output symbols $\hat{\bm{z}}=[\hat{\bm{z}}^{h_0}_1,\hat{\bm{z}}^{h_0}_2,\hat{\bm{z}}^{h_1}_3,\hat{\bm{z}}^{h_1}_4,\ldots,\hat{\bm{z}}^{h_n}_{m-1},\hat{\bm{z}}^{h_n}_m]$.
	\end{enumerate}
	
	Each dynamic encoding process can be summarized as follows:
	\begin{equation}
		\{{\bm{z}^{h_n}_i}\}^m_{i=p} = g_{\bm{\zeta}}(\{{\bm{z}^{h_0}_i}\}^m_{i=p},h_0,h_n),
	\end{equation}
	where $g$ and $\bm{\zeta}$ denote the dynamic encoder and its parameter set, respectively,
	$n$ signifies the count of changes in the channel gain coefficient,
	$p$ to $m$ denote the sequence numbers of blocks remaining untransmitted and
	$(\{{\bm{z}^{h_n}_i}\}^m_{i=p})$ denote the blocks 
	designated for the subsequent transmission stage. These parameters hold true for a range of $n$ values from $1$ onward, with $p$ taking values from $1$, $2$, and so on, up to $m$.
	
	Since the decoder is designed as the reverse structure of the encoder, we have not included the static/dynamic decoder in Fig. \ref{fig:transmission}. Each dynamic decoding process can be expressed as:
	\begin{equation}
		\{{\bm{\hat{z}}^{h_0}_i}\}^t_{i=v} = G_{\bm{\psi}}(\{{\bm{\hat{z}}^{h_n}_i}\}^t_{i=v},h_0,h_n),
	\end{equation}
	where $G$ and $\bm{\psi}$ represent the dynamic decoder and its parameter set, respectively, $v$ to $t$ denote the sequence numbers of blocks transmitted under the channel condition of $h_n$. 
	The static decoding process can be expressed as:
	\begin{equation}
		\hat{\bm{x}} = F_{\bm{\phi}}(\bm{\hat{z}}^{h_0},h_0) \in \mathbb{R}^l,
	\end{equation}
	where  
	 $F$ and $\bm{\phi}$ represent the static decoder and its parameter set, respectively, and $\bm{\hat{x}}$ denotes an estimate of the original image $\bm{x}$.
	The distortion between the original image $\bm{x}$ and the reconstructed image $\bm{\hat{x}}$ is expressed as:
	\begin{equation}
		d(\bm{x},\bm{\hat{x}}) =\frac{1}{l} ||\bm{x}-\bm{\hat{x}}||^{2}.
	\end{equation}
	Additionally, we define the channel bandwidth ratio as $R=k/l$.
	Under a certain $R$, the goal is to determine the encoder and decoder parameters $\bm{\theta}^*$, $\bm{\zeta}^*$, $\bm{\phi}^*$, and $\bm{\psi}^*$ that minimize the expected distortion as follows:
	\begin{equation}
	(\bm{\theta}^*, \bm{\zeta}^*, \bm{\phi}^*, \bm{\psi}^*) = \underset{\bm{\theta}, \bm{\zeta}, \bm{\phi}, \bm{\psi}}{\arg \min}E_{p(\bm{x},\hat{\bm{x}})}[d(\bm{x},\hat{\bm{x}})],
	\end{equation}
	where $\bm{\theta}^*$, $\bm{\zeta}^*$, $\bm{\phi}^*$, and $\bm{\psi}^*$ denote the optimal solutions for each parameter set, and $p(\bm{x},\bm{\hat{x}})$ represents the joint probability distribution of $\bm{x}$ and  $\bm{\hat{x}}$.
	\begin{figure}[t]
		\centering 
		\includegraphics[width=0.95\linewidth]{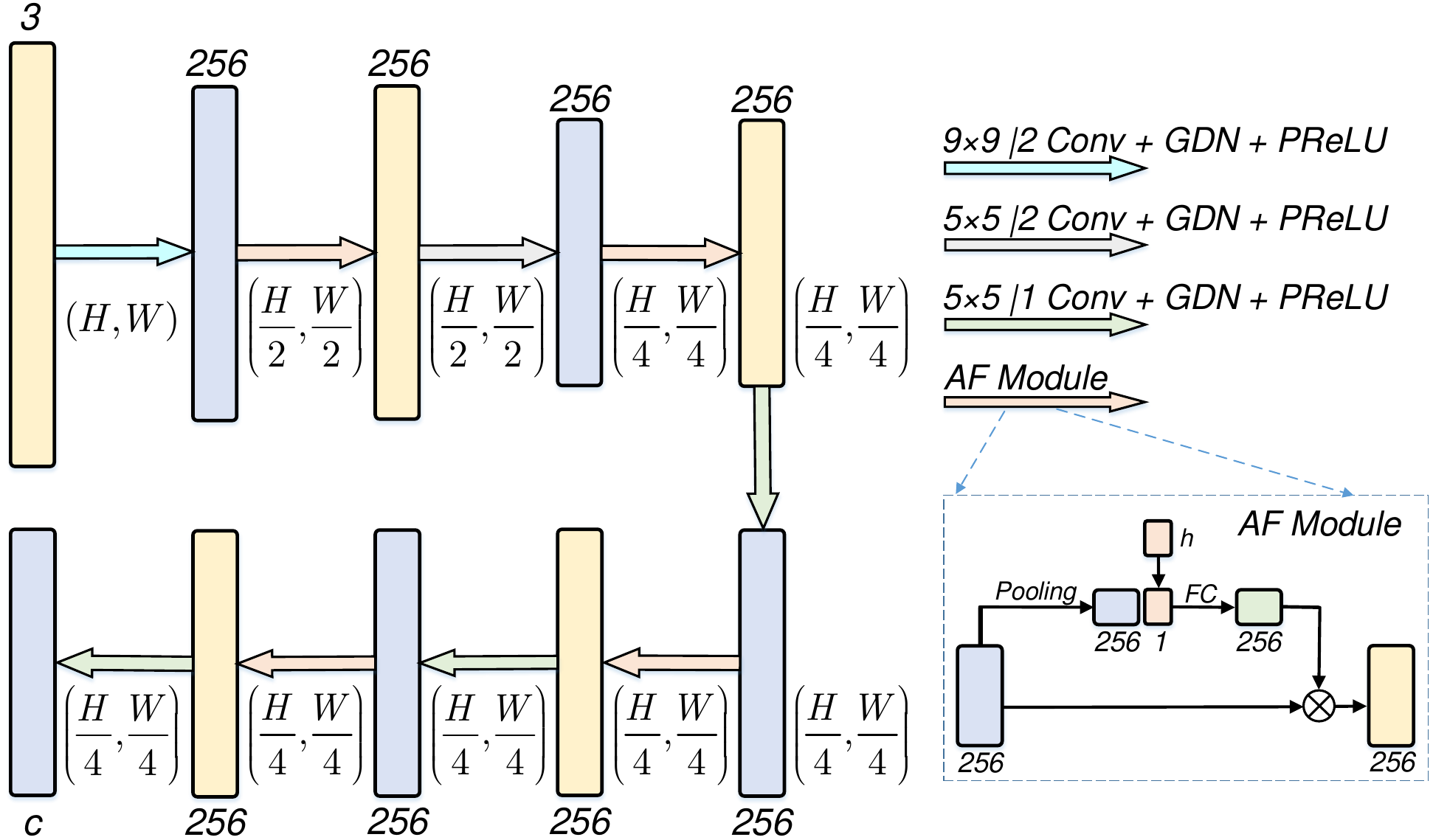} 
		\captionsetup{font={footnotesize  }}
		\captionsetup{justification=raggedright,singlelinecheck=false}
		\caption{The architecture of the IC module.}
		\label{fig:IC} 
	\end{figure}
	\section{Deep Refinement Encoders} \label{Module Design}
	
	We now introduce the proposed DRJSCC method in detail with the specific focus on the encoders, as shown in Fig. \ref{fig:system}.
	The key point of the static encoder is the initial coding (IC) module and the key point of the dynamic encoder is the re-coding (RC) module.

	\subsection{IC Module}
	The IC module undertakes the task of encoding the source image into channel symbols based on the initial channel conditions. This is accomplished through the utilization of AF modules, as detailed in \cite{ADJSCC}. To implement the IC module, we adopt the network structure depicted in  Fig. \ref{fig:IC}.
	
	The IC module consists of $5$ convolutional layers and $4$ AF modules, arranged in an interleaved fashion. The notation $F\times F|S$ represents that the kernel size is $F \times F$, and the stride is $S$.
	
	\begin{figure*}[t] 
		\centering 
		\includegraphics[width=0.81\linewidth]{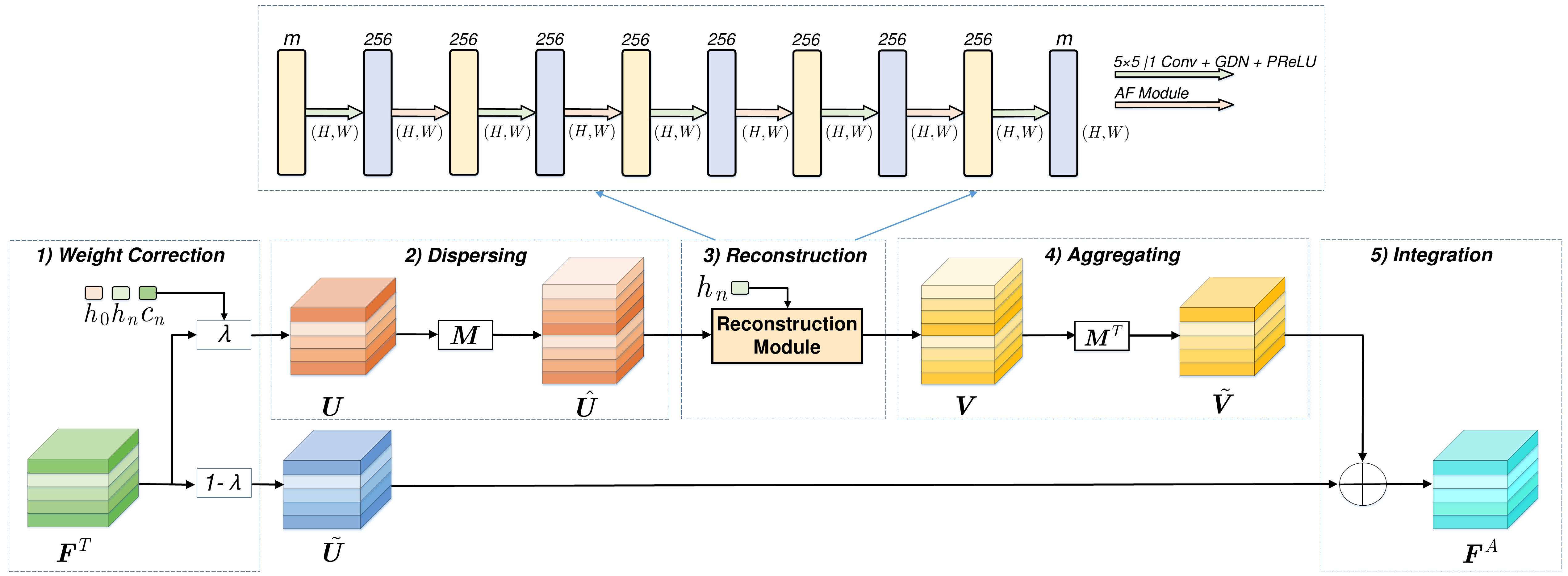} 
		\captionsetup{font={footnotesize  }}
		\captionsetup{justification=raggedright,singlelinecheck=false}
		\caption{The architecture of our proposed RC module.}
		\label{fig:RC} 
	\end{figure*}
	\subsection{RC Module}
	The RC module, as illustrated in Fig. \ref{fig:RC}, is responsible for re-encoding the remaining blocks when channel conditions undergo changes. We will provide a more detailed explanation of the various components of the RC module, which are as follows: 1) Weight Correction; 2) Dispersing; 3) Reconstruction; 4) Aggregating; 5) Integration.

	\textit{1) Weight Correction:}
	To differentiate and adapt to varying amplitudes in channel condition fluctuations, we introduce a parameter $\lambda \in (0,1)$. This parameter is used to regulate the refinement intensity, as described below:
	\begin{equation}
		\left\{
		\begin{array}{ll}
			\bm{U} = \lambda \cdot \bm{F}^T \\
			\tilde{\bm{U}} = (1-\lambda) \cdot \bm{F}^T
		\end{array}	
		\right.,
	\end{equation}
	where $\bm{F}^T$, $\bm{U}$, and $\tilde{\bm{U}}$ represent the input of the RC module, the portion to be re-encoded, and the portion that is not subject to re-encoding, respectively. Moreover, $\lambda$ can be learned by a neural network as follows:
	\begin{equation}
		\lambda = 
		\begin{cases}
			0 & |h_0| < |h_n|\\
			\mathcal{P}(h_0,h_n,c_n) & |h_0| > |h_n|
		\end{cases},	
	\end{equation}
	where $h_0$ and $h_n$ represent the initial and current channel gain coefficients, respectively, $c_n$ represents the number of blocks in $\bm{F}^T$, and $\mathcal{P}$ represents the network consisting of two fully connected (FC) layers. 
	Our experiments indicate that when the channel conditions exhibit improvement, refinement becomes unnecessary. In such instances, we set $\lambda$ to a value of $0$.

	\textit{2) Dispersing:} 
	The block number $c_n$  within  $\bm U$  introduces uncertain,  making it  challenging  to process $\bm U$ efficiently using either convolutional layers or fully connected layers. To address this challenge, we disperse the features in $\bm{U}$ into $\hat{\bm{U}}$ using a predetermined block number $m$.
	The method is similar to transposed convolution, but the convolution kernel is customized. The kernel size $K$ is $m-c_n+1$.
	The relationship between $\bm{U}$ and $\hat{\bm{U}}$ can be represented by a matrix: $\bm{U}\bm{M}=\hat{\bm{U}}$. The expression of $\bm{M}$ is shown as follows:
	\begin{equation}
		\resizebox{0.9\hsize}{!}{$
			\bm{M}=
			\left(
			\begin{matrix}
				w & w & w & \dots & w & 0 & 0 & \dots & 0 & 0 \\
				0 & w & w & \dots & w & w & 0 & \dots & 0 & 0 \\
				0 & 0 & w & \dots & w & w & w & \dots & 0 & 0 \\
				\vdots & \vdots & \vdots &  & \vdots & \vdots & \vdots &  & \vdots & \vdots \\
				0 & 0 & 0 & \dots & w & w & w & \dots & w & 0 \\
				0 & 0 & 0 & \dots & 0 & w & w & \dots & w & w \\
			\end{matrix}
			\right)_{c_n \times m},
			$}
	\end{equation}
	where the number of $w$ in each row is $K$, and $w = 1/K$.
	
	\textit{3) Reconstruction:}
	Moreover, we undertake the task of reconstructing the features in $\hat{\bm{U}}$ to obtain $\bm V$ while taking into account the current CSI. This operation can be effectively executed by making use of AF modules as described in \cite{ADJSCC}. 
	Therefore, we leverage the network structure depicted in  Fig. \ref{fig:RC}, which bears similarities to the network in the IC module. However, we adjust the parameters for the convolution layers, as the reconstruction does not compress features.
	The reconstruction process can be expressed as:
	\begin{equation}
		\bm{V}=\mathcal{S}_{\bm{\gamma}}(\hat{\bm{U}},h_n),
	\end{equation}
	where $\mathcal{S}$ and $\bm{\gamma}$ denote the reconstruction module and its parameter set, respectively, and $h_n$ denotes the current channel gain coefficient. 
	
	\textit{4) Aggregating:}
	Furthermore, we aggregate the features in $\bm{V}$ into $\tilde{\bm{V}}$ composed of $c_n$ blocks, thus ensuring that the RC module does not affect the channel bandwidth ratio $R$.
	The method is similar to convolution but the convolution kernel is customized. The kernel size $K$ is also $m-c_n+1$.
	The relationship between $\bm{V}$ and  $\tilde{\bm{V}}$ can also be simply expressed as $\bm{V}\bm{M}^T=\tilde{\bm{V}}$, where $\bm{M}^T$ is the transpose of $\bm{M}$.
	
	\begin{figure}[t] 
		\centering 
		\includegraphics[width=0.9\linewidth]{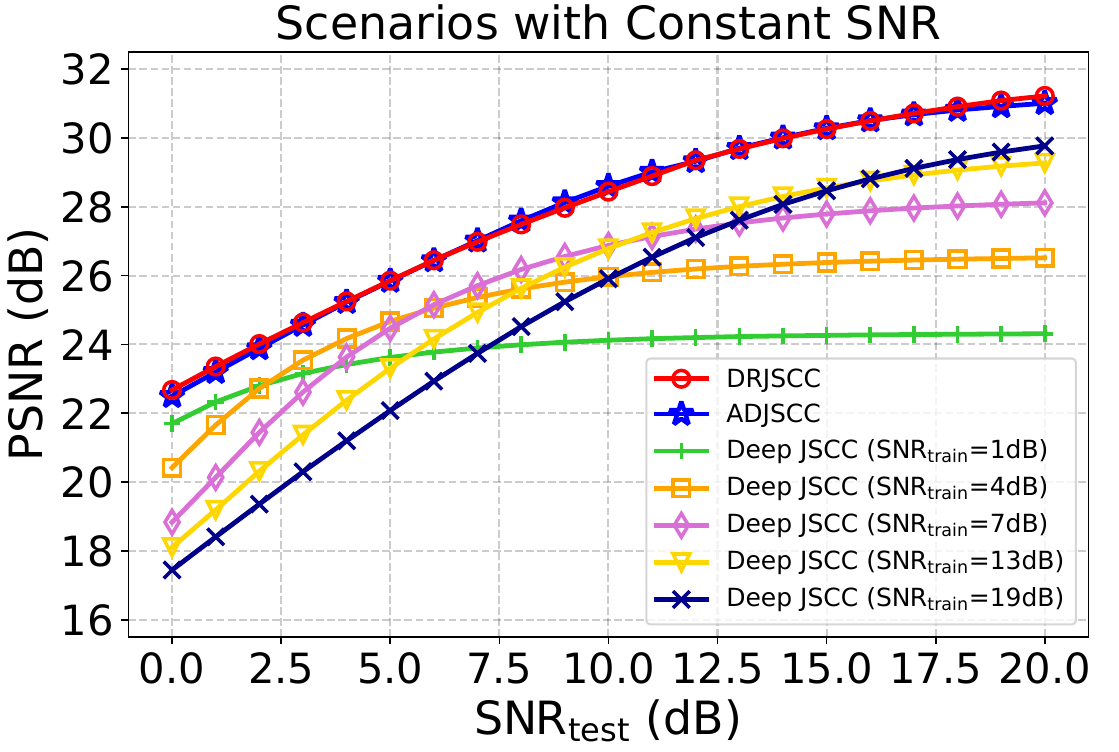} 
		\captionsetup{font={footnotesize  }}
		\captionsetup{justification=raggedright,singlelinecheck=false}
		\caption{Performance comparison of DRJSCC, ADJSCC, and deep JSCC in the scenarios with constant channel conditions. }
		\label{Fig:S1_invariable} 
	\end{figure}
	\begin{figure*}[t] 
		\centering 
		\includegraphics[width=0.91\linewidth]{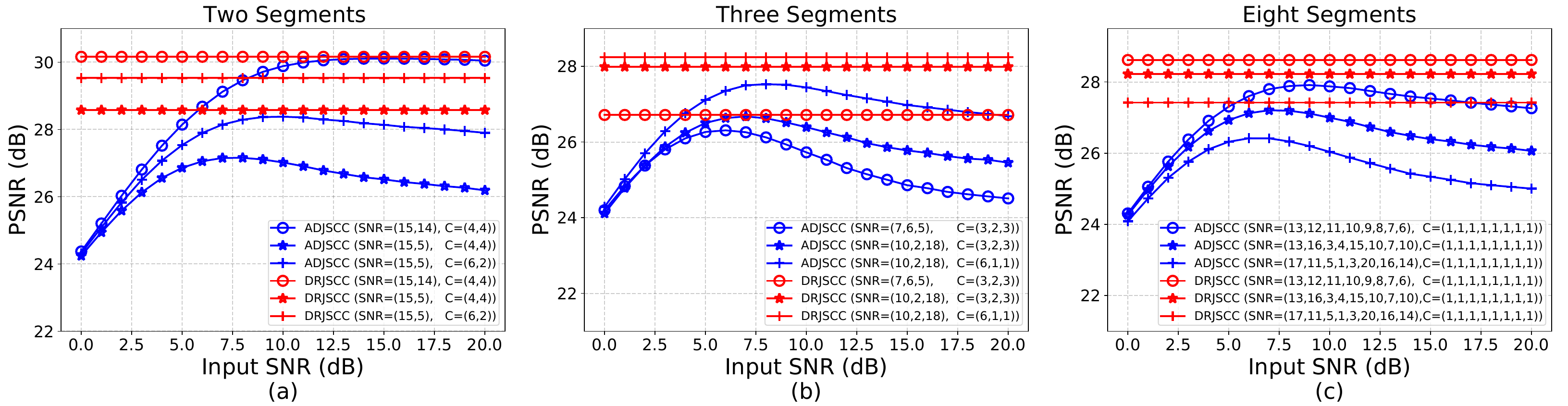} 
		\captionsetup{font={footnotesize  }}
		\captionsetup{justification=raggedright,singlelinecheck=false}
		\caption{Performance comparison between DRJSCC and ADJSCC under the SNR variation channel. (a) Scenarios where SNR changes once. (b) Scenarios where SNR changes twice. (c) Scenarios where SNR varies in each block. }
		\label{Fig:S2_variable} 
	\end{figure*}
	\textit{5) Integration:}
	Finally, inspired by \cite{ResNet}, we further integrate the features in $\tilde{\bm{V}}$ and $\tilde{\bm{U}}$ 
	to make the network form a residual-like structure,
	thereby accelerating convergence and reducing the loss introduced by excessively deep networks.
	The process can be expressed as $\bm{F}^A= \tilde{\bm{V}}+\tilde{\bm{U}}$,
	where $\bm{F}^A$, as the output of the RC module, will be transmitted over the channel after normalization.
	
	\section{Simulation Results}
	\subsection{ Simulation Setup  }
	To compare our DRJSCC model with  existing DL-based JSCC models, we implement the proposed DRJSCC using the deep learning platform ``Pytorch". 
	We employ the ``AdamW" optimizer with a learning rate of $1$ × $10^{-4}$, set the batch size to $128$, and conduct training for $1,920$ epochs.
	We use the CIFAR-$10$ dataset, which includes $50,000$  images of size $3$ × $32$ × $32$ in the training dataset and $10,000$ images in the test dataset.
	The performance is quantified in terms of the peak signal-to-noise ratio (PSNR), which is defined as follows: 
	\begin{equation}
		\textrm{PSNR}=10\log _{10}\frac{\textrm{MAX}^{2}}{\textrm{MSE}} (\textrm{dB}),
	\end{equation}
	where MSE $=d(\bm x,\bm {\hat{x}})$ denotes the mean square-error (MSE) between the source image $\bm x$ and the reconstructed image 
	$\bm{\hat{x}}$. Moreover, MAX is the maximum possible value of the pixels,  e.g., MAX equals $255$ for the images of RGB format. 
	
	\begin{figure*}[t] 
		\centering 
		\includegraphics[width=0.82\linewidth]{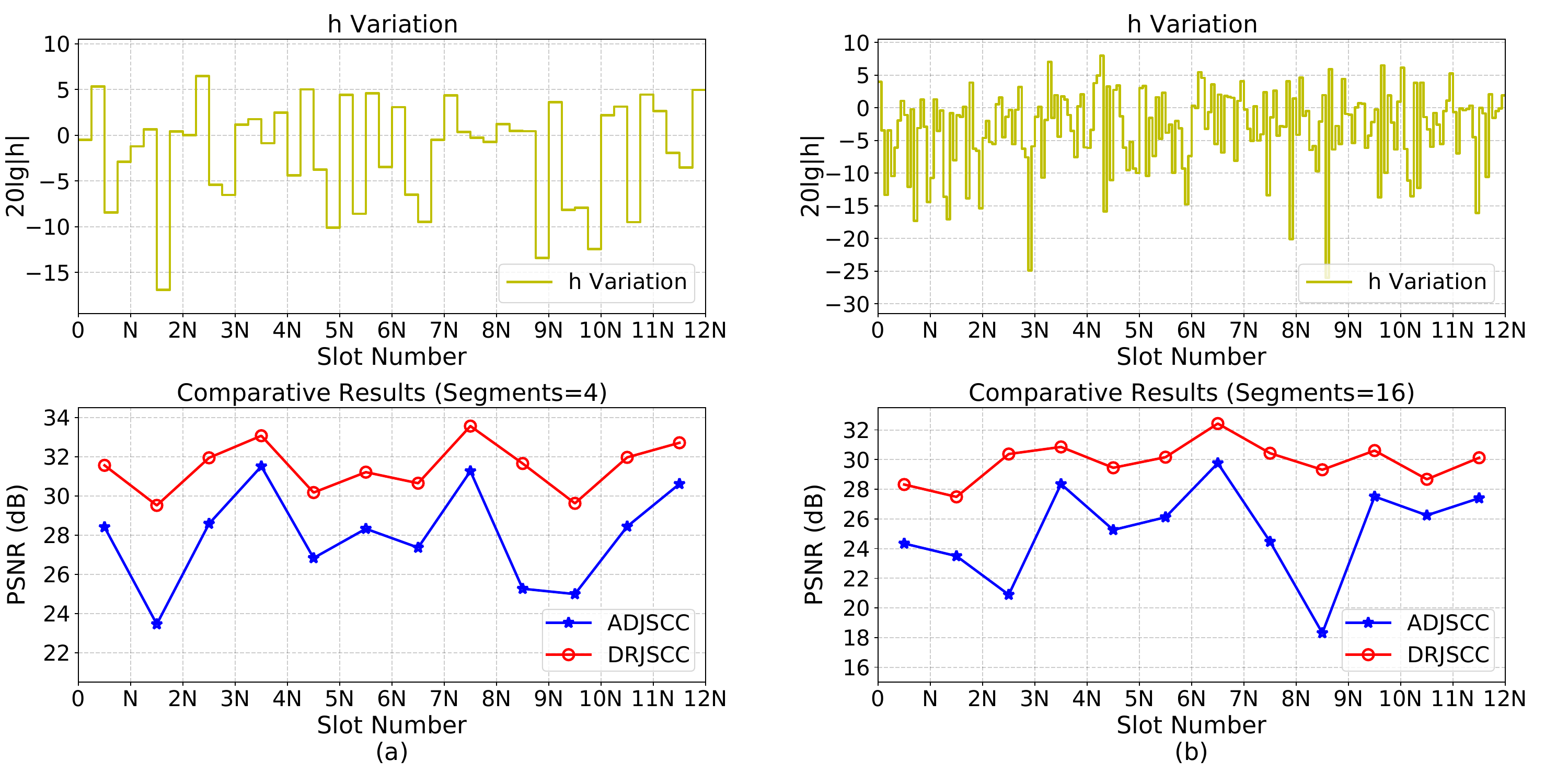} 
		\captionsetup{font={footnotesize  }}
		\captionsetup{justification=raggedright,singlelinecheck=false}
		\caption{Performance comparison between DRJSCC and ADJSCC under the Rayleigh fading channel. (a) Channel coefficient $h$ changes every $4$ blocks. (b) Channel coefficient $h$ changes every $1$ block.}
		\label{Fig:F5_12scen_ray} 
	\end{figure*}
	
	\subsection{ Simulation under the SNR Variation Channel }
	We consider that most existing DL-based JSCC models are mainly designed for a simple channel model, i.e., AWGN, in which SNR is used to characterize CSI. Therefore, for a fair comparison, we begin the simulation by temporarily ignoring the channel gain coefficient $h$ but focusing on the instantaneous SNR. At this point, we feed the real-time SNR into the encoder to achieve refinement.
	The bandwidth ratio $R$ is set to $1/12$, and we divide the encoded symbols into $8$ blocks.
	The existing DL-based JSCC models ADJSCC and deep JSCC are adopted for comparison. In the following experiments, both DRJSCC and ADJSCC are trained under a uniform SNR distribution within the range [$0$, $20$] dB.
	However, for DRJSCC, SNR changes over each block, while for ADJSCC, SNR changes every 8 blocks. All of the deep JSCC models are trained at specific $\textrm{SNR}_{\textrm{train}}$ $=$ $1$ dB, $4$ dB, $7$ dB, $13$ dB, and $19$ dB, respectively. 
	
	We initially simulate the scenarios with constant channel conditions, which will serve as the baseline for our subsequent experiments. Fig. \ref{Fig:S1_invariable} shows the comparative results. We observe that the performance of the DRJSCC model is better than any deep JSCC model trained at the specific $\textrm{SNR}_{\textrm{train}}$. Even when $\textrm{SNR}_{\textrm{train}}$ $=$ $\textrm{SNR}_{\textrm{test}}$, our DRJSCC  still outperforms the deep JSCC with a gap of approximately $1$ dB. On the other hand, compared with ADJSCC, our DRJSCC achieves a similar performance.
	In conclusion, even in the scenarios with constant channel conditions, our DRJSCC still  exhibits competitive performance  compared with the existing models.
	
	As mentioned above, the advantage of our model lies in the scenarios with time-varying channel conditions.  For simplicity, we only compare our DRJSCC with  ADJSCC in this part, because the ADJSCC model exhibits significantly better performance than any deep JSCC model. Fig. \ref{Fig:S2_variable}  compares the performance between DRJSCC and  ADJSCC under various channel conditions. In this section, we set the horizontal axis as the input SNR of ADJSCC so as to reduce the influence of channel mismatch on  ADJSCC.  $``\textrm{SNR}$ $=$ $(\textrm{S}_{1},\textrm{S}_{2}),\textrm{C}$ $=$ $(\textrm{C}_{1}, \textrm{C}_{2})"$ in the legend denotes that the $8$ blocks mentioned above are divided into $2$ segments and the first segment contains $\textrm{C}_{1}$  blocks with the channel condition of $\textrm{SNR}=\textrm{S}_{1}$ dB, but the second segment contains $\textrm{C}_{2}$ ($\textrm{C}_{2}=8-\textrm{C}_{1}$) blocks with the channel condition of $\textrm{SNR}=\textrm{S}_{2}$ dB.  
	
	Fig. \ref{Fig:S2_variable}(a) and Fig. \ref{Fig:S2_variable}(b) show the comparative results in  scenarios where SNR changes once and twice, respectively. It can be observed that when the amplitude of the SNR variation is small, the performance of  DRJSCC is similar to that of ADJSCC. However, when the amplitude of the SNR variation is large, our DRJSCC achieves much better performance than ADJSCC, even after reducing the channel mismatch of  ADJSCC. 
	Additionally, we obverse that when more blocks are transmitted over the changed channel, the performance gap between DRJSCC and ADJSCC becomes larger. This demonstrates the advantage of dynamically refining the encoding strategy, 
	ensuring that it consistently adapts to the time-varying channel conditions.
	Fig. \ref{Fig:S2_variable}(c) shows the results in the scenario where SNR varies in each block, which can be related to situations where the code length is extremely long or the coherence time is extremely short. In this case, the performance of  ADJSCC consistently falls behind that of  DRJSCC since the input SNR of  ADJSCC can no longer effectively characterize the channel conditions at this point. As a result, the encoding strategy is not the optimal. In contrast, our DRJSCC customizes the encoding strategy for each block at this time and  maintains a high transmission performance.

	\subsection{Simulation under the Rayleigh Fading Channel  }
	It is more practical to consider scenarios with varying channel conditions under the Rayleigh fading channel. Therefore, we further compare DRJSCC with ADJSCC in this case. 
	 We set $R$  to $1/6$ and divide the encoded symbols into $16$ blocks. To more intuitively show the refinement effect achieved by DRJSCC, we consider $12$ scenarios that adhere to the characteristics of the Rayleigh fading channel 
	and evaluate our DRJSCC on these  scenarios. Specifically, we fix $\sigma^2$ to $10$ dB. Building upon this, we randomly generate $h$  according to complex Gaussian distribution to set up $12$ scenarios. Furthermore, $128$ randomly selected images will be transmitted through DRJSCC and ADJSCC in each of these $12$ scenarios, and the average PSNR of these $128$ images will be used as the score of DRJSCC and ADJSCC.
	
	Fig. \ref{Fig:F5_12scen_ray} displays the comparative results between DRJSCC and ADJSCC.
	Each interval of $N$ represents a scenario.
	Fig. \ref{Fig:F5_12scen_ray}(a) shows the results when $h$ changes every $4$ blocks and Fig. \ref{Fig:F5_12scen_ray}(b) shows the results when $h$ changes every $1$ block. 
	From a holistic perspective, it is clear that in the context of the Rayleigh fading channel, our DRJSCC outperforms ADJSCC by a significant margin.
	This is because the Rayleigh fading channel enlarges the amplitude of the channel condition variation, making the refinement effect of DRJSCC more prominent.
	From a local point of view, we observe that as long as one block of the encoded channel symbols is transmitted under extremely bad channel conditions, the ADJSCC model experiences substantial performance degradation, even if the other blocks are minimally affected by noise. In contrast, our DRJSCC handles this situation more flexibly. 
	Additionally, comparing Fig. \ref{Fig:F5_12scen_ray}(a) with Fig. \ref{Fig:F5_12scen_ray}(b), we can observe that the coherence time has a great impact on the performance of ADJSCC. In scenarios with a long coherence time, as shown in Fig. \ref{Fig:F5_12scen_ray}(a), ADJSCC can maintain a moderate performance. However, in scenarios with a short coherence time, as shown in Fig. \ref{Fig:F5_12scen_ray}(b), the performance of ADJSCC deteriorates. On the contrary, our model exhibits strong adaptability to  coherence time and consistently maintains a better performance, which  arises from its ability to dynamically refine the encoding strategy based on real-time channel conditions, thereby minimizing the adverse effects of each channel condition variation.
	
	\section{Conclusion} \label{Conclusion}
	In our study, we introduced an innovative method known as DRJSCC, designed to seamlessly adapt to time-varying channel conditions. The approach involves the division of channel symbols into discrete blocks, facilitating a progressive transmission process. When channel conditions undergo shifts, the method dynamically refines the encoding strategy by re-encoding the remaining blocks, ensuring alignment with the prevailing channel state. Our simulations demonstrated that our DRJSCC delivered comparable performance to existing approaches in stable channel conditions while showing superior robustness against time-varying channels. This adaptability makes it a versatile choice for a wide range of scenarios.
	\bibliographystyle{IEEEtran}           
	\bibliography{IEEEabrv,Reference}      

\begin{thebibliography}{10}
\providecommand{\url}[1]{#1}
\csname url@samestyle\endcsname
\providecommand{\newblock}{\relax}
\providecommand{\bibinfo}[2]{#2}
\providecommand{\BIBentrySTDinterwordspacing}{\spaceskip=0pt\relax}
\providecommand{\BIBentryALTinterwordstretchfactor}{4}
\providecommand{\BIBentryALTinterwordspacing}{\spaceskip=\fontdimen2\font plus
\BIBentryALTinterwordstretchfactor\fontdimen3\font minus
  \fontdimen4\font\relax}
\providecommand{\BIBforeignlanguage}[2]{{%
\expandafter\ifx\csname l@#1\endcsname\relax
\typeout{** WARNING: IEEEtran.bst: No hyphenation pattern has been}%
\typeout{** loaded for the language `#1'. Using the pattern for}%
\typeout{** the default language instead.}%
\else
\language=\csname l@#1\endcsname
\fi
#2}}
\providecommand{\BIBdecl}{\relax}
\BIBdecl

\bibitem{Shannon}
C.~E. Shannon, ``{A mathematical theory of communication},'' \emph{Bell Syst.
  Tech. J.}, vol.~27, no.~3, pp. 379--423, Jul. 1948.

\bibitem{JPEG}
C.~Christopoulos, A.~Skodras, and T.~Ebrahimi, ``{The JPEG2000 still image
  coding system: An overview},'' \emph{IEEE Trans. Consum. Electron.}, vol.~46,
  no.~4, pp. 1103--1127, Nov. 2000.

\bibitem{Turbo}
C.~Berrou, A.~Glavieux, and P.~Thitimajshima, ``{Near shannon limit
  error-correcting coding and decoding: Turbo-codes. 1},'' in \emph{Proc. IEEE
  Int. Conf. Commun. (ICC)}, vol.~2, May 1993, pp. 1064--1070.

\bibitem{JSCC_better}
V.~Kostina and S.~Verdú, ``{Lossy joint source-channel coding in the finite
  blocklength regime},'' \emph{IEEE Trans. Inf. Theory}, vol.~59, no.~5, pp.
  2545--2575, May 2013.

\bibitem{compare}
P.~Jiang, C.-K. Wen, S.~Jin, and G.~Y. Li, ``{Deep source-channel coding for
  sentence semantic transmission with HARQ},'' \emph{IEEE Trans. Commun.},
  vol.~70, no.~8, pp. 5225--5240, Aug. 2022.

\bibitem{DeepSC}
H.~Xie, Z.~Qin, G.~Y. Li, and B.~Juang, ``Deep learning enabled semantic
  communication systems,'' \emph{IEEE Trans. Signal Process.}, vol.~69, pp.
  2663--2675, Apr. 2021.

\bibitem{speech}
Z.~Weng and Z.~Qin, ``Semantic communication systems for speech transmission,''
  \emph{IEEE J. Sel. Areas Commun.}, vol.~39, no.~8, pp. 2434--2444, Aug. 2021.

\bibitem{video}
T.-Y. Tung and D.~Gündüz, ``{DeepWiVe: Deep-learning-aided wireless video
  transmission},'' \emph{IEEE J. Sel. Areas Commun.}, vol.~40, no.~9, pp.
  2570--2583, Sep. 2022.

\bibitem{DeepJSCC}
E.~Bourtsoulatze, D.~Burth~Kurka, and D.~Gündüz, ``{Deep joint source-channel
  coding for wireless image transmission},'' \emph{IEEE Trans. Cogn. Commun.
  Netw.}, vol.~5, no.~3, pp. 567--579, Sep. 2019.

\bibitem{Deep_F}
D.~B. Kurka and D.~Gündüz, ``{DeepJSCC-f: Deep joint source-channel coding of
  images with feedback},'' \emph{IEEE J. Sel. Areas Inf. Theory}, vol.~1,
  no.~1, pp. 178--193, May 2020.

\bibitem{UDeepSC}
G.~Zhang, Q.~Hu, Z.~Qin, Y.~Cai, G.~Yu, and X.~Tao, ``A unified multi-task
  semantic communication system for multimodal data,'' \emph{arXiv preprint
  arXiv:2209.07689}, 2022.

\bibitem{ADJSCC}
J.~Xu, B.~Ai, W.~Chen, A.~Yang, P.~Sun, and M.~Rodrigues, ``{Wireless image
  transmission using deep source channel coding with attention modules},''
  \emph{IEEE Trans. Circuits Syst. Video Technol.}, vol.~32, no.~4, pp.
  2315--2328, Apr. 2022.

\bibitem{Deep_V}
W.~Zhang, H.~Zhang, H.~Ma, H.~Shao, N.~Wang, and V.~C.~M. Leung, ``{Predictive
  and adaptive deep coding for wireless image transmission in semantic
  communication},'' \emph{IEEE Trans. Wireless Commun.}, vol.~22, no.~8, pp.
  5486--5501, Aug. 2023.

\bibitem{ResNet}
K.~He, X.~Zhang, S.~Ren, and J.~Sun, ``{Deep residual learning for image
  recognition},'' in \emph{Proc. IEEE Conf. Comput. Vis. Pattern Recog.
  (CVPR)}, Jun. 2016, pp. 770--778.

\end{thebibliography}

\end{document}